\documentclass[usenatbib,letter]{mnras} 
\usepackage{graphicx, amssymb, natbib, bm}
\usepackage{xcolor}

\usepackage{mathptmx}
\usepackage{epsfig}
\usepackage{wrapfig}
\usepackage{amsmath}
\usepackage{natbib}
\usepackage{graphicx}
\usepackage{subfigure}
\usepackage{enumitem}
\usepackage{breqn}

\newcommand\msun{\hbox{\ensuremath{{\rm M}_{\odot}}}}
\newcommand\sfrha{SFR\ensuremath{_{{\rm H}\alpha}}}
\newcommand\sfruv{SFR\ensuremath{_{\rm UV}}}
\newcommand\aha{A\ensuremath{_{{\rm H}\alpha}}}

\def\lsun{\hbox{\ensuremath{{\rm L}_{\odot}}}}
\def\ha{\hbox{\ensuremath{{\rm H}\alpha}}}

\def\dhst{3D-HST}
\def\mperyr{\hbox{\ensuremath{\rm \msun~yr^{-1}}}}

\newcommand{\scinot}[1]{\ensuremath{\times 10^{#1}}}

\def\ntwo{\hbox{[N\sc{ii}]}}

\newcommand{\cse}{Computing in Science \& Engineering}

\hypersetup{draft}
\title[Bursty SFHs in Massive Transition Galaxies]{Evidence for Non-smooth Quenching in Massive Galaxies at \boldmath{$z\sim1$ }}
\author[Carleton et al.]
{Timothy Carleton$^{1}$\thanks{$\!\!$e-mail: carletont@missouri.edu},
	Yicheng Guo$^1$, Hooshang Nayyeri$^2$, Michael Cooper$^2$,
	\newauthor Gregory Rudnick$^3$, Katherine Whitaker$^{4,~5}$ \\
	$\!\!^{1}$Department of Physics and Astronomy, 223 Physics Building, University of Missouri, Columbia, MO 65211, USA \\
	$\!\!^2$Center for Cosmology, Department of Physics and Astronomy, 
	4129 Reines Hall, University of California, Irvine, CA 92697,
	USA \\
	$\!\!^3$The University of Kansas, Department of Physics and Astronomy, Malott Room 1082, 1251 Wescoe Hall Drive, Lawrence, KS, 66045, USA \\
	$\!\!^4$Department of Physics, University of Connecticut, Storrs, CT  06269, USA \\
	$\!\!^5$Department of Astronomy, University of Massachusetts, 710 North Pleasant St, LGRT-524, Amherst, MA 01003, USA
}

\begin{document}
	
	\pagerange{\pageref{firstpage}--\pageref{lastpage}} 
	\pubyear{2019}
	
	\maketitle
	
	\begin{abstract}
		We investigate a large sample of massive galaxies at $z\sim1$ with combined {\it HST} broad-band and grism observations to constrain the star-formation histories of these systems as they transition from a star-forming state to quiescence. 
		Among our sample of massive ($M_*>10^{10}~\msun{}$) galaxies at $0.7<z<1.2$, dust-corrected \ha{} and UV star-formation indicators agree with a small dispersion ($\sim0.2$~dex) for galaxies on the main sequence, but diverge and exhibit substantial scatter ($\sim0.7$~dex) once they drop significantly below the star-forming main sequence. 
		Significant \ha{} emission is present in galaxies with low dust-corrected UV SFR values as well as galaxies classified as quiescent using the $UVJ$ diagram.
		We compare the observed \ha{} flux distribution to the expected distribution assuming bursty or smooth star-formation histories, and find that massive galaxies at $z\sim1$ are most consistent with a quick, bursty quenching process.
		This suggests that mechanisms such as feedback, stochastic gas flows, and minor mergers continue to induce low-level bursty star formation in massive galaxies at moderate redshift, even as they quench.
	\end{abstract}


	\section{Introduction}
	A growing consensus of observations indicates that the population of massive quiescent galaxies has been building up since before $z=3$ \citep{bell2004,faber2007,ilbert2013}. The presence of this population at early epochs poses a significant challenge to our current understanding of galaxy formation and evolution, as these systems must have formed early and suffered a quick shutdown in star formation \citep[`quenching';][]{peng2010,thomas2010,kuntschner2010,vandokkum2015,daddi2005,goddard2017}. Many mechanisms have been proposed to cause this shutdown of star formation in massive galaxies, such as the build-up of a hot-gas halo \citep{keres2005,dekel2006}, feedback from an Active-Galactic-Nucleus \citep[AGN;][]{dimatteo2005,hopkins2006,beckmann2017} or star formation activity \citep{oppenheimer2006,ceverino2009} driven by a recent merger \citep{hopkins2014}, or the stabilization of the cold gas against fragmentation \citep{martig2009}. However, the importance/feasibility of these processes, and how they may evolve with redshift, is still uncertain and remains a major unanswered question in our current understanding of galaxy evolution.

	As many authors have noted \citep{martin2007,schawinski2014,wild2016,belfiore2016,pandya2017}, understanding of the processes involved in quenching can be discerned through detailed studies of massive galaxies in the process of transitioning from star-forming to quiescent. Determining the star-formation histories of these galaxies can constrain which processes drive quenching. For example, \cite{schawinski2007} found that local early-type galaxies are consistent with a short ($<250$~Myr) quenching associated with merger-driven feedback, but late-type galaxies are consistent with a longer (many Gyr) quenching process such the buildup of a hot gas halo through radio-mode AGN feedback \citep{croton2006}. 
	Similarly, by using semi-analytic models describing the star-formation histories of massive galaxies, \cite{pandya2017} found that a fast quenching mode is the predominant quenching mode at high $z$, whereas low-$z$ galaxies are associated with a slower quenching process. At $z=1.4-2.6$, \cite{zick2018} investigate the spectra of transition galaxies at, finding that quenching occurs on a $100-200$~Myr timescale, and using photometric data of galaxies between $z=0.25$ and $z=3.75$, \cite{carnall2018} find that most massive galaxies are consistent with quenching times $\lesssim 1$~Gyr.
	Studies investigating the abundance of galaxies with SFRs in between the star-forming and quiescent populations over cosmic time find that quenching takes place on $\sim1-3$~Gyr timescales \citep{wetzel2013,balogh2004,hahn2017}.
	
	However, the photometric signatures relied upon by these studies are predominantly sensitive to B and A stars tracing the average SFR in the past few hundred Myr --- any $10-100$~Myr variations don't leave an imprint on them \citep{worthey1997}. A growing body of evidence suggests that star formation in low-mass galaxies is dominated by episodes of bursty star-formation activity where the instantaneous SFR can vary by nearly an order of magnitude on $\sim 10$~Myr timescales \citep{guo2016,weisz2012,sparre2017}. Whether massive galaxies experience this same level of burstiness as they quench can help elucidate the processes at play as their star-formation activity shuts down  \citep{french2018}.

	This bursty star-formation activity is usually identified through the ratio of \ha{} and UV star-formation indicators. Nebular emission from H{\sc ii} regions around O stars lasting $\sim10$~Myr closely traces the immediate SFR, whereas UV emission from B and A stars lasting $\sim200$~Myr traces the average SFR over longer timescales. In depth studies have shown that \ha{} and UV tracers agree for galaxies with ongoing star formation at a level above $0.1$~\mperyr{} in the local Universe \citep{salim2007,fumagalli2011} and above $10$~\mperyr{} at higher redshifts \citep[][but see \citealt{wisnioski2019}]{reddy2010,shivaei2015}. Deviations from this agreement have been used as evidence for bursty star-formation activity in dwarf galaxies \citep{guo2016,weisz2012}. Indeed, simulations of star-formation in low-mass galaxies find that the \ha{}/UV luminosity ratio varies in a way consistent with the assumed bursty nature of star formation \citep{sparre2017}. However, these results are generally limited to low-mass galaxies with high specific SFRs. It is unclear if \emph{high-mass} galaxies with \emph{low specific SFRs} show this behavior as well.
	
	In this paper, we present evidence of bursty star formation in massive transition galaxies (galaxies more than $1$~dex below the main sequence, but not completely quenched) at $z\sim1$, suggesting that these galaxies experience a bursty decline in star-formation activity, rather than a smooth one. In section~\ref{sec:data}, we describe our sample selection and SFR measurements. In section~\ref{sec:models}, we compare the observed \ha{} and UV SFRs and describe the model SFHs used in our analysis. In section~\ref{sec:results} we compare our measurements with the predictions of the model SFHs, and in section~\ref{sec:systematics} we describe possible systematic effects on our results. Section~\ref{sec:conclusions} summarizes our conclusions. Throughout this study, we assume a $\Lambda$CDM cosmology with H$_0=70$~km~s$^{-1}$~Mpc$^{-1}$, $\Omega_{\rm m}=0.3$, and $\Omega_\Lambda=0.7$. Except for when otherwise indicated, we assume a \cite{chabrier2003} IMF.
	
	\begin{figure*}
		\centering
		\includegraphics[width=1\linewidth]{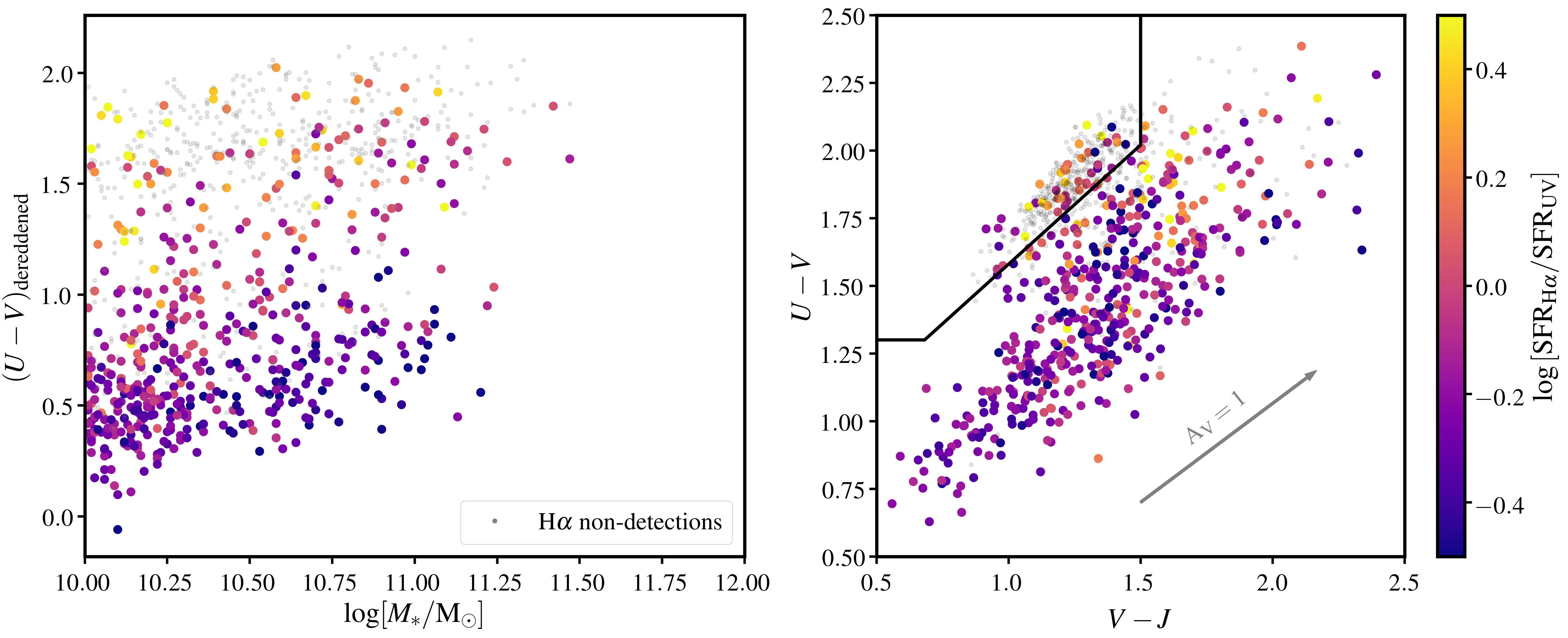}
		\caption{Our sample in both $M_*$-vs-dereddened $U-V$ space (left) and the $UVJ$-diagram (right). For objects with \ha{} detections, the color of the point corresponds the the ratio between \ha{} and UV SFRs. While \ha{}-detected objects are primarily blue, continuously star-forming objects, there is a significant population of red, quiescent galaxies with \ha{} SFRs comparable to their UV SFRs.}
		\label{fig:uvjhauv}
	\end{figure*}
	
	\begin{figure}
		\centering
		\includegraphics[width=1\linewidth]{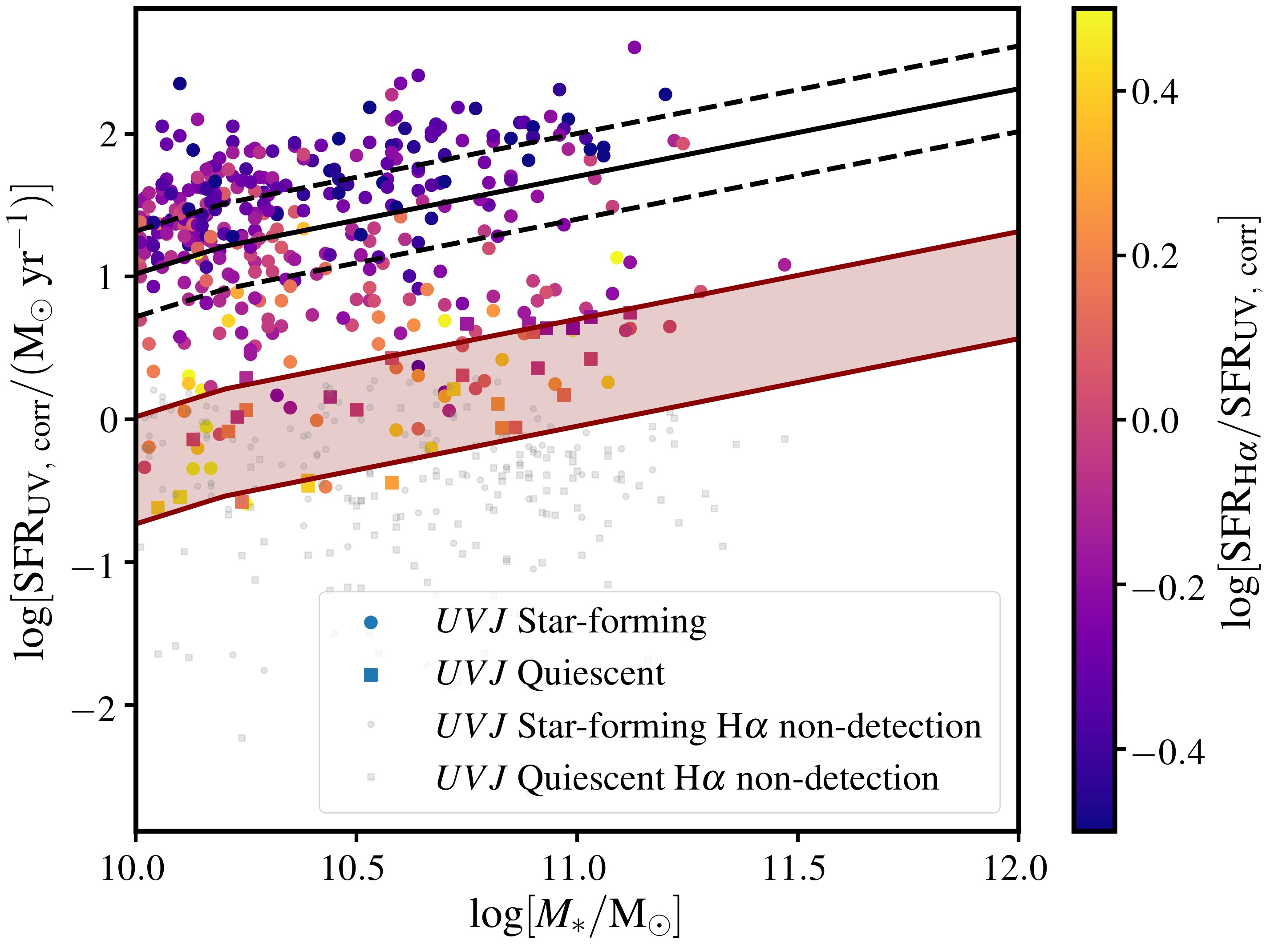}
		\caption{Our sample in SFR-$M_*$ space. As in Figure~\ref{fig:uvjhauv}, objects with \ha{} emission are color-coded by the ratio of their \ha{} and UV SFRs and grey points show objects with \ha{} non-detections. Points and squares show objects classified as $UVJ$ star-forming and $UVJ$-quiescent respectively. The black solid and dashed lines show the main sequence from \protect\citep{whitaker2014} and $0.3$~dex scatter respectively. The red shaded region corresponds to between $1$ and $1.75$~dex below the main sequence, where we focus our investigation. While the ratio of UV-to-\ha{} SFR is uniform for objects on the main sequence, there is a substantial variation for objects below the main sequence.}
		\label{fig:mainsequenceplot}
	\end{figure}

	\section{Data}
	\label{sec:data}
	Our data is primarily drawn from the \dhst{} survey \citep{skelton2014,momcheva2016}. This survey targets the CANDELS fields  \citep{grogin2011,koekemoer2011} with the G141 grism, which covers $1.1$ to $1.65$ $\mu{\rm m}$ and traces \ha{} emission between $z=0.7$ to $z=1.5$. We make use of stellar masses, rest-frame colors, \ha{} fluxes, and redshifts published in the \dhst{} catalogs. The derivation of these parameters is described in detail in \cite{momcheva2016}. Below, we briefly summarize these calculations.

	First, accompanying $JH_{\rm F140W}$ and $H_{\rm F160W}$ direct-exposure images have been used to identify all
	objects in the \dhst{} footprint with $JH_{\rm F140W}<26$. Images were reduced with the {\sc calwf3}
	package, and grism spectra were extracted utilizing the {\sc aXe} pipeline, using direct exposure images
	for source extraction and contamination estimation. Photometry was carried out on the
	direct-exposure images and combined with publicly-available optical and near-IR photometry to
	create observed spectral energy distributions (SEDs). These SEDs were fit with template SEDs to
	measure the photometric redshift \citep[using EAZY;][]{brammer2008} and theoretical SEDs to determine the stellar mass
	\citep[using FAST;][]{kriek2009}. Rest-frame colors were determined by fixing the template redshift at the best-fit redshift ({\sc z\_best})
	and refitting the SED to the photometry, only using observed filters, $i$, for which $|\lambda_{{\rm
			obs},i} - \lambda_{{\rm rest},j}|<1000$~\AA{} and measuring the flux through the rest-frame filter $j$.
	
	In our sample we select from the $1754$ galaxies for which the \dhst{} catalogs contain a measurement of the \ha{} flux, have a stellar mass (M$_*$) above $10^{10}$~\msun{}, and are between $z$ of $0.7$ and $1.2$. These limits are identified so that \ha{} is detectable well below the main sequence: the $3\sigma$ \ha{} detection limit taken from \cite{momcheva2016} reaches $1.3$~dex below the main sequence at $z=1.2$ for point sources with no extinction. Additionally, we make the following cuts to our sample:
	\begin{enumerate}[leftmargin=\parindent,labelwidth=\parindent]
		\item We exclude $112$ X-ray detected AGN identified in the CANDELS catalogs \citep{ueda2008,salvato2011,xue2011,rangel2013,nandra2015} from our sample.
		\item Because accurate rest-frame UV luminosities, which are derived from the UV portion of the best-fit SED, are critical to this analysis, we further restrict our sample to objects with good photometry (as determined by the  {\sc use\_phot} flag in the \dhst{} catalog), excluding $97$ objects. We also exclude $486$ predominantly star-forming objects where the reduced $\chi^2$ of the best-fit SED is greater than $2$. Many objects with poor SED fits either have a nearby companion or a disturbed morphology, suggesting that inconsistent aperture photometry is the cause of the high $\chi^2$ values.
		\item To ensure accurate \ha{} measurements, $14$ objects are excluded for which the grism coverage is incomplete within $100$~\AA{} of \ha{} at the best-fit redshift. 
		\item To avoid spurious \ha{} measurements, we exclude $2$ objects for which the contamination level is more than $50\%$ of the total flux and $47$ for which the contamination at the wavelength of \ha{} is $>50\%$.
		\item For galaxies whose dust-corrected UV SFRs (see Sec.~\ref{sec:sfrmeasure}) are at least $1$~dex below the main sequence, we visually inspect both the 1D and 2D grism spectra to verify that the emission is not due to contamination, a bad redshift, or any anomalies in the spectrum, removing $17$ additional objects.
		\item Because constraints on the level of extinction from SED fitting in the CANDELS catalogs are necessary to accurately correct our measurements for dust extinction, we exclude the $216$ objects in the GOODS-N field, as the CANDELS-based SED-fitting in that field is not complete.
	\end{enumerate}
	These cuts leave $780$ galaxies overall, $417$ of which have \ha{} emission above the $3\sigma$ level. All objects have at least one observation blue-ward of rest-frame $2800$~\AA{}, and $88\%$ of objects have least one detection in that wavelength range, so the NUV luminosity is well constrained by observations.

	Figure~\ref{fig:uvjhauv} illustrates this sample in the  $UVJ$ diagram, as well as a diagram showing the sample in $M_*$ vs. dereddened $(U-V)$ space.  The observed $U-V$ colors are dereddened using a \cite{calzetti2000} extinction law and extinction from the best-fit SED in the CANDELS catalogs (see Sec.~\ref{sec:sfrmeasure}). Figure~\ref{fig:mainsequenceplot} shows the sample in SFR-$M_*$ space.
	In both figures, points are color-coded by the ratio between \ha{} and UV SFRs for objects with \ha{} emission. It is already clear that, while most objects with \ha{} emission are classified as star-forming, $11\%$ of $UVJ$-quiescent objects and $20\%$ of objects more than $1$~dex below the main sequence have significant \ha{} emission. To specifically investigate the nature of the SFH of galaxies in the process of quenching, we narrow our focus further to the $312$ systems ($92$ of which are $UVJ$-star forming and $220$ of which are $UVJ$-quiescent) whose dust-correct UV SFRs are between $1$ and $1.75$~dex below the main sequence of \cite{whitaker2014} in our modeling (Section~\ref{sec:models}). This space is highlighted in Figure~\ref{fig:mainsequenceplot}.

	\subsection{SFR measurements}
	\label{sec:sfrmeasure}
	The UV-based SFR, which is sensitive to stars less than $100-200$~Myr old, is derived from the UV luminosity at $2800$~\AA{} following the \cite{wuyts2011} conversion:
	\begin{equation}
	\label{eqn:uvsfr}
	{\rm log}\left({\rm SFR_{UV}}\right)={\rm log}\left(L_{2800}\right)-9.44+0.4 A_{\rm UV},
	\end{equation}
	where \sfruv{} is the dust-corrected UV SFR in units of \mperyr{} and $L_{2800}$ is the luminosity at $2800$~\AA{} in units of \lsun{}  taken from the best-fit SED. We assume a \cite{calzetti2000} extinction law, such that $A_{\rm UV}=7.26 E(B-V)$, where $E(B-V)$ is the $V$-band reddening. Because low-SFR galaxies (at least $1$~dex below the main sequence) in our analysis have relatively little dust ($E(B-V)\sim0.1$), the adoption of an alternative extinction law has a negligible effect on our results. For example, adopting an SMC law \citep{gordon2003} alters the UV SFRs by $<0.1$~dex and the ratio between UV and \ha{} SFRs by $<0.1$~dex for $90\%$ of low-SFR objects. Deep imaging from $HST$, the CFHT, and Subaru telescope sample this region of the SED to a $3\sigma$ depth of approximately $27.5$, resulting in an unobscured SFR limit of $\sim0.3$~\mperyr{} at $z=1.2$.
	
	At high redshifts, a majority of the UV light from star-formation is absorbed and re-emitted in the far-IR \citep{whitaker2014}. While far-IR measurements from $Herschel$ or ground-based sub-mm telescopes can measure this directly for a limited number of bright objects, most survey-based studies rely on a luminosity-dependent conversion from $Spitzer$ $24$\micron{} flux to a total IR luminosity \citep{whitaker2014,whitaker2017}. However, diffuse dust heated from old stars, in addition to emission directly from asymptotic-giant branch (AGB) stars, can contribute substantially to the observed $24$~$\mu$m flux for galaxies with low SFRs \citep{piovan2003,marigo2008,kelson2010,fumagalli2014}. Additionally the conversion between polycyclic aromatic hydrocarbon (PAH) emission (the origin of $24$~\micron{} emission at high $z$) and SFR depends on the age and ionizing flux of the stellar population \citep{shivaei2017}, which may vary significantly across our sample.
	
	\begin{figure}
		\centering
		\includegraphics[width=1\linewidth]{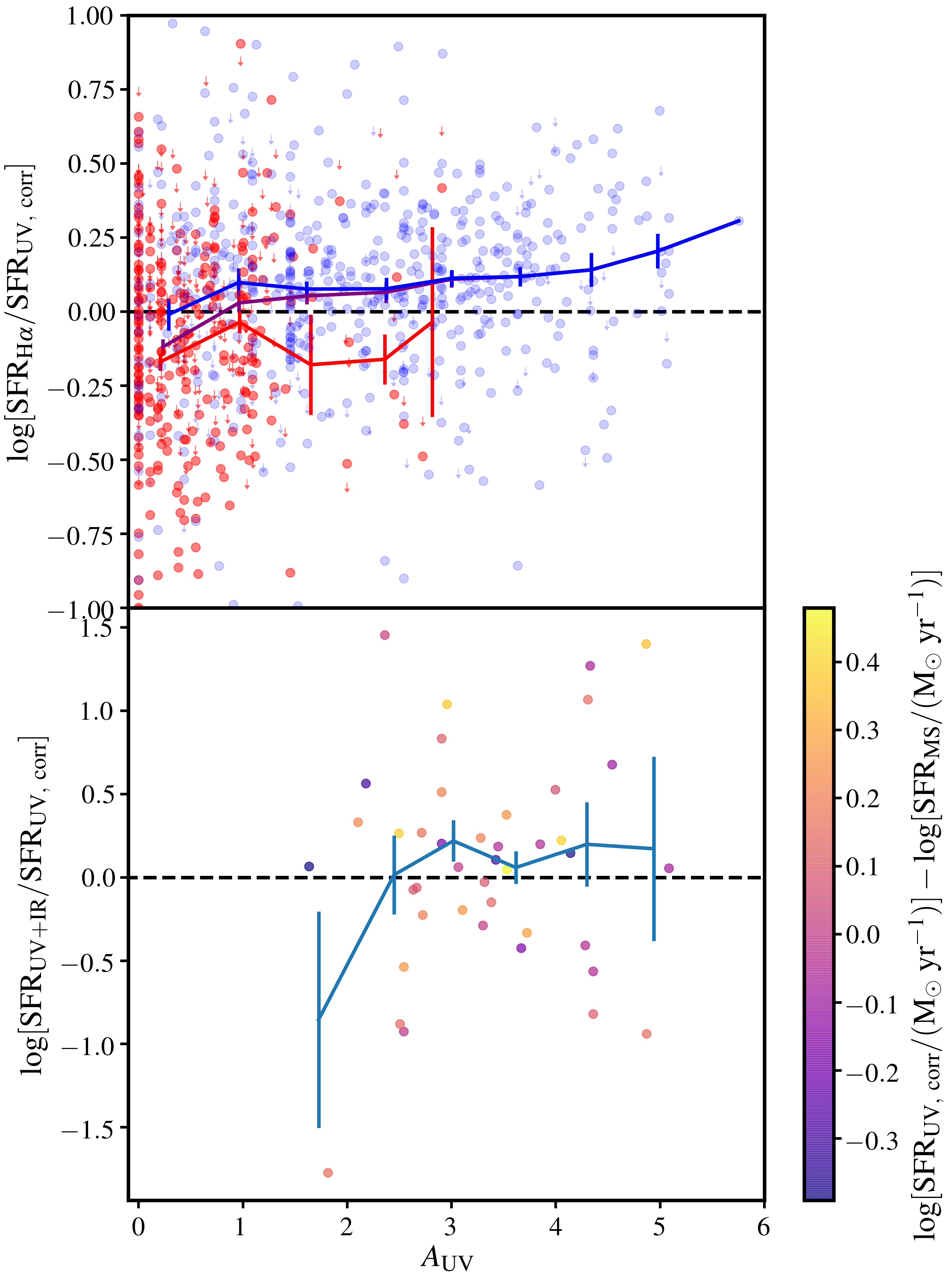}
		\caption{Verification of our extinction measurements. {\bf Top:} The upper plot shows the ratio of \ha{}-to-UV SFR as a function of $A_{\rm UV}$, with $UVJ$-quiescent objects shown as red points and $UVJ$-star forming objects shown as blue points. Although $UVJ$-quiescent objects have lower \ha{}/UV SFRs than $UVJ$-star-forming objects (see Sec.~\ref{sec:results}), there is no correlation between extinction and \ha{}/UV SFR. This implies that our assumed ratio of nebular-to-continuum extinction is not affecting our \ha{} SFR values.
		{\bf Bottom:} The ratio UV+IR SFRs with {\it Herschel} detections to our dust-corrected UV SFRs. Objects in this plot are further restricted to objects whose UV-corrected SFRs are $>20$~\mperyr{} to avoid selection effects. Points are color-coded by offset from the main sequence. The SFR measurements agree and there is no correlation with A$_{\rm UV}$ or distance from the main sequence, suggesting that our dust-corrected UV SFRs are accurate.}
		\label{fig:irvsuv}
	\end{figure}

	Given these uncertainties, we elect to use the UV luminosity corrected for extinction, which primarily derives from young stars, to measure the star-formation activity on these timescales. By comparing our dust-corrected UV SFRs with {\it Herschel}-based SFRs available for a subset of our sample ($142$ objects, $42$ of which have dust-corrected UV SFRs greater than $20$~\mperyr{}), we have determined that using the extinction reported in the \dhst{} catalogs results in a correlation between extinction and the ratio of {\it Herschel}-based SFRs-to-dust-corrected UV SFRs, even among galaxies for which {\it Herschel} observations are complete.
	Alternatively, using the median $E(B-V)$ reported in the CANDELS catalogs, combining $14$ different SED fits with different assumptions \citep{nayyeri2017,stefanon2017,guo2013,galametz2013}, results in agreement between the UV-corrected SFRs and UV+IR SFRs within $50\%$ and independent of the amount of extinction for objects with UV-corrected SFRs greater than $20$~\mperyr{} (see Fig.~\ref{fig:irvsuv}). Moreover, Balmer-decrement-based extinction measurements for objects in the LEGA-C survey \citep{vanderwel2016} agree with the CANDELS extinction measurement better than the \dhst{} measurement. The UV-corrected SFRs agree with the SFR of the best-fit SED for star-forming objects, further suggesting that the dust-corrected UV SFRs are accurate.
	For objects with low SFRs, the above recipe may overestimate their true SFRs because emission from Post-AGB stars, Blue-Horizontal-Branch Stars, and Blue Stragglers can represent a non-negligible fraction of the UV luminosity \citep{dorman1995}. Because this emission depends on the SFH of the galaxy in a non-trivial way, we incorporate it into our modeling (see Sec.~\ref{sec:models} and \ref{sec:uvevolved}) rather than subtracting this emission when calculating the UV SFR.

	The \ha{}-based SFRs (\sfrha{}) are calculated from the \cite{kennicutt2012} conversion adjusted to a Chabrier IMF following \cite{muzzin2010}: 
	\begin{equation}
	\label{eqn:hasfr}
	{\rm log}\left({\rm SFR}_{{\rm H} \alpha}\right)={\rm log}\left(L_{{\rm H} \alpha}\right)-7.77+0.4 A_{{\rm H} \alpha},
	\end{equation}
	where $L_{{\rm H}\alpha}$ is the luminosity of the \ha{} line in solar luminosities, and \aha{} is the internal extinction of the \ha{} line. The $3\sigma$ H$\alpha$ flux limit achieved by \dhst{} of $2.1\scinot{-17}$ erg/s/cm$^2$ corresponds to a H$\alpha$ SFR of $0.8$~\mperyr{} at $z=1.2$ assuming $\aha{}=0$.
	To accurately measure \ha{} SFRs, we correct the \ha{} luminosity for (1) extinction, (2) stellar absorption, (3) emission from post-AGB stars, and (4) contamination from nearby \ntwo{} emission (in that order).
	
	The \ha{} SFRs are corrected for extinction following a \cite{calzetti2000} law. Following \cite{wuyts2013}, we relate the nebular extinction to the continuum emission as $\aha{}=1.9A_{\rm cont}-0.15A_{\rm cont}^2$, where $A_{\rm cont}$ is the continuum extinction at \ha{}. For a subsample ($13$) of objects with H$\beta$ detections from the LEGA-C survey \citep{vanderwel2016}, the Balmer-decrement-based extinction measurements are generally consistent with the extinction determined from the continuum extinction, with an median deviation of $0.32$~mag and dispersion of $1.2$~mag.
	Among these objects, there is no correlation between the deviation and the measured extinction.

	We correct for \ha{} absorption, which can be significant for massive galaxies \citep{kauffmann2003}, using an age-dependent factor based on the amount of absorption in spectra generated following the model SFHs used in our analysis (see Sec.~\ref{sec:models}). On average, for our star-formation histories, this varies with age according to: 
	\begin{equation}
	\begin{array}{lcr}
	\log\left(\ha{}~{\rm EW}\right)=&&\\
	\begin{cases}
	0.431 &  t_{\rm lw}<6\scinot{8}~{\rm Gyr} \\
	-0.11 \left(\log t_{\rm LW}\right)^2+0.18 \left(\log t_{\rm LW}\right) 
	+0.53   &  t_{\rm lw}>6\scinot{8}~{\rm Gyr} 
	\end{cases}
	\end{array}
	\label{eqn:haabs}
	\end{equation}
	where $\ha{}$~EW is the \ha{} is the equivalent width of absorption and $t_{\rm LW}$ is the bolometric light-weighted age of the stellar population in Gyr. For each galaxy, the light-weighted age from the the \dhst{} catalog is used in conjunction with Equation~\ref{eqn:haabs} to determine the amount of \ha{} absorption. However, if a constant absorption of $3$~\AA{} is adopted, the change in our results is negligible. For star-forming galaxies, this correction lowers the sSFR by $\sim10^{-10.3}$~yr$^{-1}$; for older quiescent galaxies, it corresponds to a decrease of $\sim10^{-10.8}$~yr$^{-1}$.
	
	Furthermore, post-AGB stars can produce enough ionizing radiation to contribute significantly to the \ha{} luminosity \citep{cidfernandes2011,singh2013,belfiore2016}.
	Although there remains uncertainty with regard to the specifics of AGB and post-AGB stellar evolution, models generally agree that evolved stars provide an ionizing flux of $\sim 10^{41}$ photons$/$s$/$\msun{} \citep{cidfernandes2011} independent of age. Assuming Case-B recombination and a temperature of $10,000$~K, this corresponds to a \ha{} luminosity per stellar mass of $1.37\scinot{29}$~erg~s$^{-1}$~\msun{}$^{-1}$. Given that evolved stellar-populations have \ntwo{}/\ha{} ratios close to $1$ \citep{belfiore2016}, we subtract $2\times1.37\scinot{29}\times (M_*/\msun{})$~erg~s$^{-1}$ (corresponding to a sSFR of $1.2\scinot{-12}$~yr$^{-1}$) from the \ha{} luminosity to isolate the \ha{} emission associated with young stars.
	
	Because of the low spectral resolution of the grism, the measured \ha{} flux contains emission from both \ha{} and nearby \ntwo{}. To correct for this contamination, we adopt a mass-dependent correction motivated by the mass-metallicity relation. The gas-phase metallicity is estimated from the measured stellar mass assuming the redshift-dependent mass-metallicity relation of \cite{zahid2014}, and the metallicity is converted to a \ntwo{}/\ha{} flux ratio following \cite{kewley2008}. The \ha{} flux reported in the \dhst{} catalog is reduced by this ratio to determine the \ha{} flux. This physically-motivated correction (typically around $\sim25\%$ for our sample) is somewhat larger than the $20\%$ usually assumed \citep{wuyts2011}.

	\section{Results}
	\label{sec:models}
\begin{figure}
		\centering
		\includegraphics[width=1\linewidth]{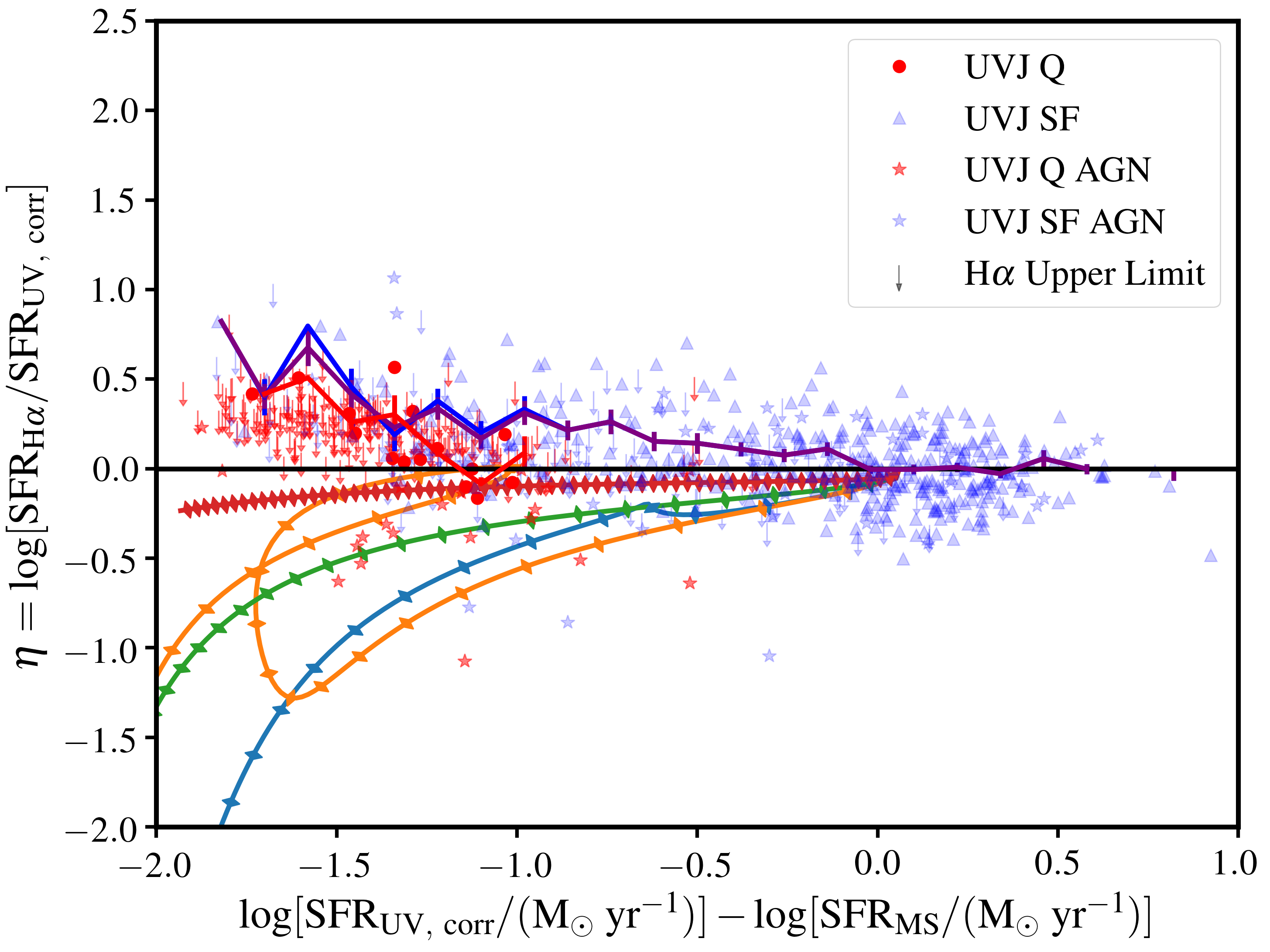}
		\caption{The relationship between $\eta$ and $\Delta$MS for galaxies in our sample. Galaxies are color-coded by whether they are classified as star-forming (blue) or quiescent (red) based on the $UVJ$ diagram. The blue, red, and purple lines show the binned relationship between $\eta$ and $\Delta$MS for $UVJ$-star-forming, $UVJ$-quiescent, and all objects with \ha{} detections respectively. Arrows illustrate the $3\sigma$ limits of galaxies without a significant \ha{} detection. Although we do not include X-ray detected AGN in our primary sample, we show them here as stars to illustrate their distribution in this space. Star-forming galaxies with AGN actually have similar $\eta$ values compared with galaxies without significant AGN, whereas quiescent galaxies with AGN have slightly less \ha{} emission than galaxies without AGN. Solid ticked lines illustrate how a galaxy following different  star-formation histories would evolve in this space, with ticks every $75$~Myr (see Sec.~\ref{sec:models}). In agreement with previous studies, \ha{} and UV SFRs match closely for galaxies on the main sequence. However, a substantial amount of scatter is present for galaxies below the main sequence. This scatter can only be reproduced with a bursty star-formation history.}
		\label{fig:ratios}
	\end{figure}

	Figure~\ref{fig:ratios} shows the ratio between \ha{} and UV SFR measurements (which we refer to as  $\eta=\log[{\rm SFR}_{{\rm H}\alpha}/{\rm \sfruv{}}])$ as a function of the offset between the UV SFR and the \cite{whitaker2014} main sequence $(\Delta$MS$)$, color coded by their location in $UVJ$ space \citep{wuyts2007,williams2009} using the \cite{whitaker2012} definition. As expected, both SFR measures agree for galaxies with ongoing star formation. As galaxies drop below the main sequence, more systems have low or undetected \ha{} emission as the instantaneous SFR (traced by \ha{}) decreases more quickly than the average SFR (traced by UV). However, there remains a significant population of systems with $\eta$ close to $0$. Notably, $11\%$ of $UVJ$-quiescent objects and $20\%$ of objects more than $1$~dex below the main sequence have significant \ha{} emission.
	
	While $UVJ$-quiescent objects with \ha{} emission have higher $E(B-V)$ values than $UVJ$-quiescent objects on average, most are characterized by $E(B-V)<0.1$, suggesting that they are generally not dusty contaminants.
	Significant $24\mu$m emission is present in only $24\%$ of $UVJ$-quiescent objects and $48\%$ of objects more than $1$~dex below the main sequence.
	Although there remains uncertainty regarding the amount of $24\mu$m emission that originates from old stars, the $24\mu$m luminosities of objects with $24\mu$m emission can generally be accounted for with a combination of low level star formation (consistent with their dust-corrected UV SFRs) and emission from an old stellar population \citep{leroy2012,salim2007,kelson2010}.
	Altogether, although dusty contaminants may be present in our sample, they likely don't represent a significant source of contamination for our study.
	
	Galaxies on the main sequence are consistent with $\eta=0$ and have a small ($\sim0.2$~dex) scatter in $\eta$, but the distribution of \ha{} fluxes and non-detections among galaxies below the main sequence implies evolution of $\eta$ as systems fall off of the main sequence. Assuming $\eta$ is normally distributed, the mean and standard deviation of that distribution that best-fit the distribution of \ha{} fluxes and non-detections among low-SFR galaxies is $-0.8$ and $0.7$~dex respectively.
	
	To address the possibility that this \ha{} emission is from bursty star formation, we illustrate how $\eta$ evolves as a function of $\Delta$MS for various model star-formation histories (SFHs). 
	All models are based on an exponentially declining plus exponential burst star-formation history following a period of constant SFR of the form:
	\begin{equation}
	{\rm SFR}(t)={\rm SFR_0} e^{-(t-t_q)/\tau_0} + \delta e^{-|t-t_q-t_1|/\tau_{\rm burst}},
	\label{eqn:sfh}
	\end{equation}
	where SFR$_0$ is the SFR of the galaxy before quenching occurs, $t_q$ is the time when the galaxy quenches, $\tau_0$ is the quenching timescale, $\delta$ is the peak burst amplitude, $t_1$ corresponds to when the burst occurs after the initial quenching, and $\tau_{\rm burst}$ is the characteristic timescale of the burst.
	First, we consider two bursty models with $\tau_{\rm burst}=\tau_0=100$~Myr and $\delta=1~\mperyr{}$:  models A and B have $t_1=0.3$~Gyr and $t_1=1.25$~Gyr respectively. 
	Although $\tau_0$ is not constrained in general, our choice of $\tau_0\sim100$~Myr is motivated by studies of recently quenched galaxies finding stellar ages consistent with short timescales \citep{zick2018,french2018,belli2019}. Additionally, we consider two no-burst models to compare with our bursty models. Model C is a smooth model with $\tau_0=200$~Myr and closely resembles model A in most other aspects. Model D, which is characterized by $\tau_0=1$~Gyr, represents the null hypothesis of quenching too slow to alter the \ha{}/UV SFR ratio (models with $\tau_0$ longer than the $\sim200$~Myr lifetime of a B star quickly resemble the $1$~Gyr model). These models are summarized in Table~\ref{tab:modeltable}. For each SFH, we model the stellar population using the {\sc pyfsps} code \citep{conroy2009,conroy2010,foremanmackey2014}.
	We initialize all models with $5$~Gyr of continuous star formation at $10$~\mperyr{} (mimicking the formation of a $\sim10^{10}~\msun{}$ galaxy to match the initial colors and UV luminosities), after which point, SFR$(t)$ follows Equation~\ref{eqn:sfh}. The models all have a solar metallicity and a Chabrier IMF. 
	The adoption of a higher metallicty could change the inferred UV SFR, but the $\eta$ distribution of systems with $10^{10}\le M_*<10^{10.75}$~\msun{} is statistically identical to the $\eta$ distribution of systems with $M_*\ge10^{10.75}~$\msun{} among transition population (systems $1-1.75$~dex below the main sequence). This evidence, combined with the fact that stellar metallicity is observed to vary by less than $0.2$~dex across our the mass range of our sample for low-SFR galaxies \citep{choi2014,estrada-carpenter2019}, suggests that the adoption of a uniform solar metallicity for our models is not unrealistic.
	From the synthesized spectra, the UV and \ha{} SFRs are calculated according to equations~\ref{eqn:uvsfr} and \ref{eqn:hasfr}, respectively. These models are not meant to span the entire range of plausible scenarios; rather they give a general sampling of what different quenching models predict.
	
	These star-formation histories are summarized in Figure~\ref{fig:burstmodels}. As seen in Figure~\ref{fig:ratios}, $\eta$ decreases dramatically with increasing $\Delta$MS for all models with $\tau_0\le 200$~Myr. For $\tau_0=1$~Gyr, $\eta$ remains roughly constant during the quenching process. The bursty models are distinguished by a sharp increase in $\eta$ during the burst due to \ha{} emission from young stars. In particular, bursty SFHs predict a large range of $\eta$ for low SFRs present in the data, while smooth SFHs predict a narrow range of $\eta$ values at a given UV SFR.

	\begin{table}
		\centering
		\begin{tabular}{|c||c|c|c|c|}
			\hline 
			Model & $\delta$ (\mperyr{}) & $t_1$ (Gyr) & $\tau_0$ (Myr) & $\tau_{\rm burst}$ (Myr) \\ 
			\hline
			A \label{model:a} & 1 & 0.3 & 100 & 100 \\ 
			\hline 
			B \label{model:b} & 1 & 1.25 & 100 & 100 \\ 
			\hline 
			C \label{model:c} & 0 & -- & 200 & -- \\ 
			\hline 
			D \label{model:d} & 0 & -- & 1000 & -- \\ 
			\hline 
		\end{tabular} 
		\caption{The parameters of the model SFHs used in our analysis.}
		\label{tab:modeltable}
	\end{table}
	
	\section{Modeling the Transition Population}
	To test which SFH model best matches the observations, we model the expected \ha{} flux distribution among galaxies between $1$ and $1.75$~dex below the main sequence.
	\begin{figure}
		\centering
		\includegraphics[width=1\linewidth]{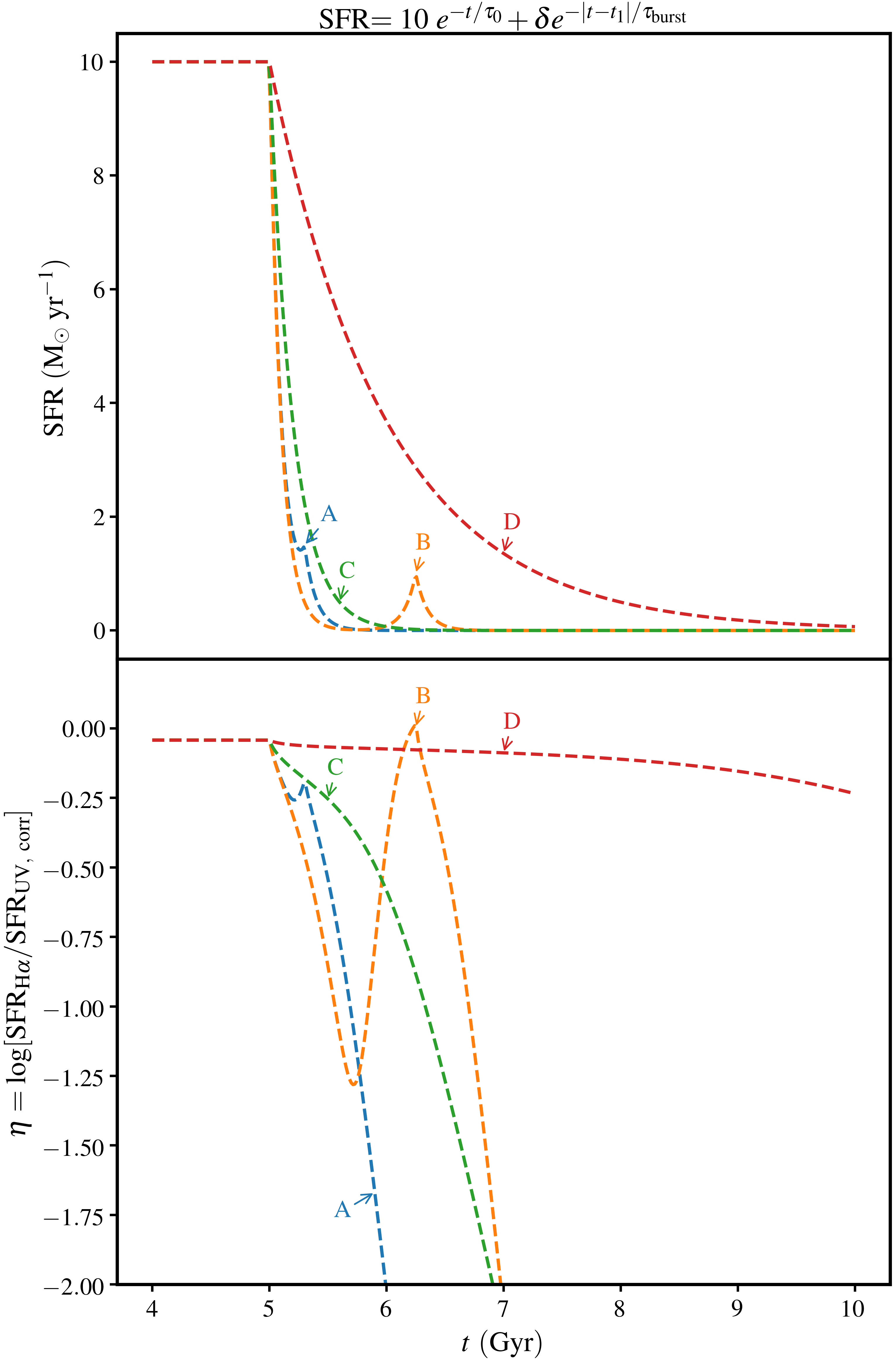}
		\caption{An illustration of the model SFHs used in this analysis. {\bf Top:} The SFR as a function of time for our models. Models A, B, and C experience a rapid decline in star formation activity. Bursty models A and B experience exponential bursts of $1$~\mperyr{}, with both the quenching and the burst characterized by $100$~Myr timescales. Models C and D describe a smooth quenching process with $\tau_0=200$~Myr and $\tau_0=1$~Gyr respectively. {\bf Bottom:} The ratio of \ha{}-to-UV SFR indicators as a function of time for our models. For slow quenching (model D), $\eta$ does not evolve substantially, whereas the rapid shutdown of star-formation in models A-C results in a correspondingly rapid decrease in \ha{} emission. This, accompanied by the slower decrease of UV emission results in quickly decreasing $\eta$ values. However, during bursts the \ha{} emission quickly rejuvenates.}
		\label{fig:burstmodels}
	\end{figure}
	For each model SFH described above, we construct an expected \ha{} flux distribution based on the objects in our sample. Specifically, we determine the expected \ha{} flux for each object given its UV SFR and the $\eta$ value predicted by each particular SFH as follows:
\begin{enumerate}[leftmargin=\parindent,labelwidth=\parindent]
	\item For each galaxy in our sample, the initial SFR (SFR$_0$) is taken to be the SFR of a galaxy with the same stellar mass and redshift on the \cite{whitaker2014} main sequence plus normally-distributed scatter of $0.3$~dex.
	\item The value of $\eta$ is estimated from the assumed SFH and the observed UV SFR. The ratio between the observed UV SFR and SFR$_0$ is taken to be SFR/SFR$_0$ in equation~\ref{eqn:sfh} and used to determine $t$, which is in turn used to determine $\eta$ (see Fig.~\ref{fig:burstmodels}). In the case that this results in multiple values of $\eta$, the dereddened diagonal $UVJ$ color ($C_{\rm SED}=0.82(U-V)-0.57(V-J)$, from \citealt{fang2018}) of the galaxy is compared with the model $C_{\rm SED}$ at the various times. The time with the closest color to the observed galaxy is used. The \ha{} SFR is determined from the $\eta$ value and the dust-corrected UV SFR.
	\item The \ha{} SFR is converted to an \ha{} luminosity.
	\item Absorption at \ha{} is subtracted from the \ha{} luminosity following Equation~\ref{eqn:haabs} using the bolometric light-weighted age as $t_{\rm LW}$.
	\item Emission from AGB stars is added to the \ha{} luminosity as $2 \times 1.37\scinot{29}$~erg~s$^{-1}$~\msun{}$^{-1}$ times the observed stellar mass (the factor of $2$ accounts for \ntwo{} emission from AGB stars, which is characterized by an \ntwo{}/\ha{} ratio of $1$).
	\item The \ha{} luminosity is corrected for attenuation and \ntwo{} contamination with the same prescriptions as described in Section~\ref{sec:data}.
	\item This luminosity is converted to flux, and a normally-distributed error of $\sigma=8\scinot{-18}$~erg/s/cm$^2$ is added to this measurement \citep{momcheva2016}.
\end{enumerate}

Although X-ray-detected AGN are excluded from our sample, we include the effects of any X-ray-non-detected AGN in our model. Using the mass and redshift-dependent AGN luminosity functions of \cite{aird2012}, we predict the fraction of our subsample expected to host AGN. We take the difference between the expected AGN occurrence and the number of observed AGN as the number of X-ray-non-detected AGN. This number of galaxies is randomly selected from our sample, and for each supposed non-detected AGN, we replace the \ha{} luminosity expected from the SFH model with the \ha{} luminosity expected given the \ha{}/UV ratio of a randomly chosen X-ray-\emph{detected} AGN.
Undetected AGN represent $10\%$ of our subsample and their \ha{}/UV ratios are not substantially different than the non-AGN population, so this correction does not have a substantial impact on the analysis. Additionally, excluding AGN based on their IRAC colors, which is more sensitive to extremely dust-extincted AGN compared with X-ray-AGN \citep{stern2005}, does not affect our conclusions.

\begin{figure}
	\centering
	\includegraphics[width=1\linewidth]{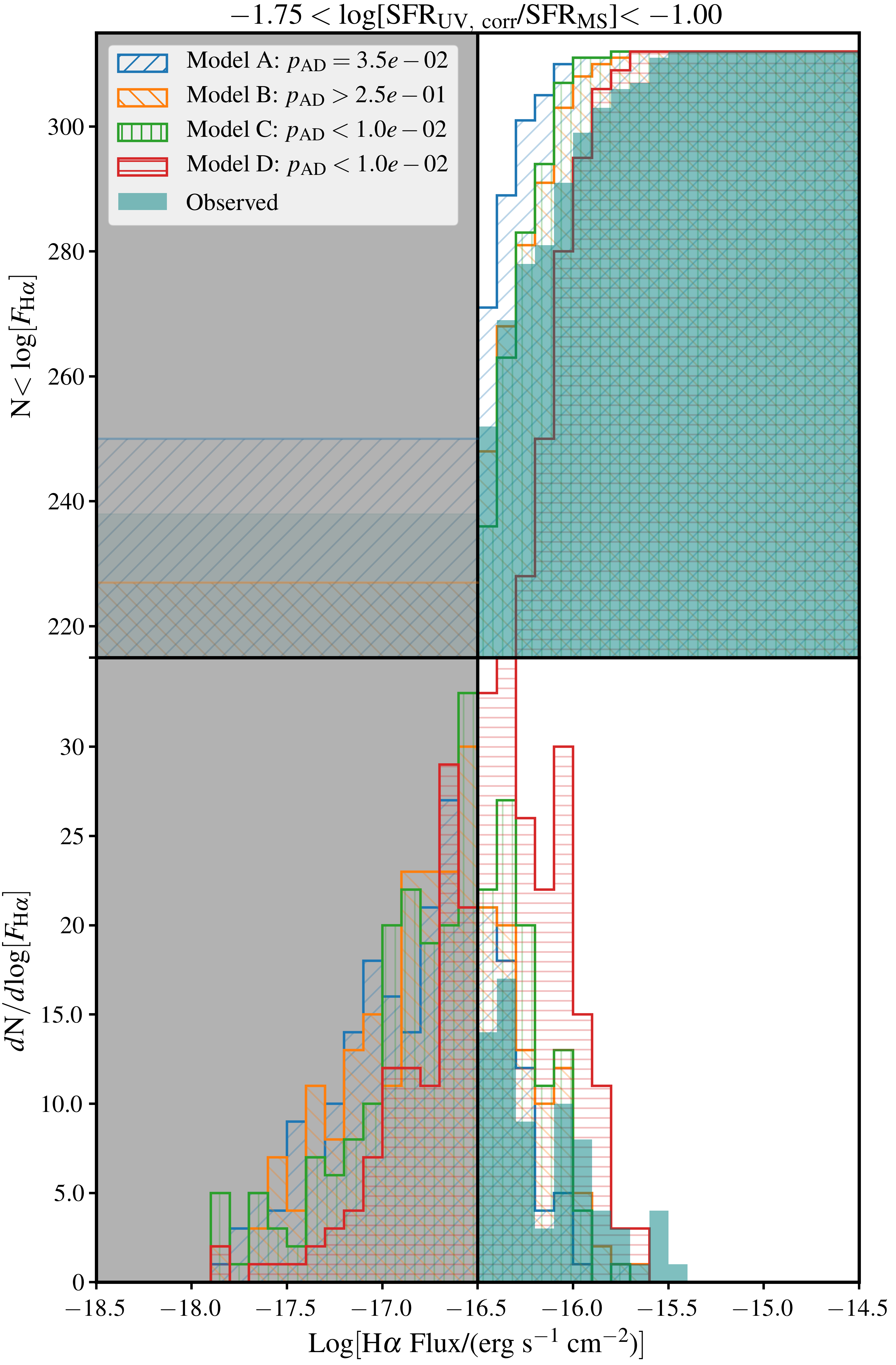}
	\caption{{\bf Bottom:} The expected \ha{} flux distribution (hatched distributions) for our various SFHs compared with observations (solid distribution) for objects with UV SFRs between $1$ and $1.75$~dex below the main sequence. For clarity, non-detections, which represent the majority of the sample, are not shown. Objects with \ha{} flux below $10^{-16.5}$~erg~s$^{-1}$~cm$^{-2}$ (shaded region) are also considered non-detections. {\bf Top:} The corresponding cumulative distributions. In the legend, we show the Anderson-Darling $p$ values comparing the theoretical distribution with the observed one (capped at $0.01$ and $0.25$). The bursty model B best represents the observed distribution. On the other hand, the smooth quenching model (model C), and the model with a burst shortly after quenching (model A) do not produce enough \ha{} emission to match the observations, and the slow quenching model (model D) produces too many \ha{} detections compared with the observations.}
	\label{fig:modelflux}
\end{figure}

	\label{sec:results}
	Figure~\ref{fig:modelflux} illustrates the distributions of \ha{}+\ntwo{} flux for both the observations (solid histogram) and the models (hatched histograms) for galaxies between $1$ and $1.75$~dex below the main sequence. Also shown are the results of an Anderson Darling test comparing the predicted \ha{} distributions with observations. For this test, all objects with \ha{} flux less than $10^{-16.5}$~erg~s$^{-1}$~cm$^{-2}$ are considered non-detections and considered to have $0$ flux.
	The slow quenching model (model D) substantially overpredicts the number of \ha{} detections. On the other hand, while the faster smooth model (model C) only slightly underpredicts the number of \ha{} detections, it significantly underpredicts the observed flux for these objects. The quick burst of model A means that by the time the UV SFR is below the main sequence, $\eta$ values quickly and unformly fall, such that it underpredicts the \ha{} flux throughout this sample. The large degree of variation in $\eta$ in model B, however, is able to reproduce the large variation in \ha{} fluxes apparent in our sample.

	\subsection{Rest-Frame Colors}
\begin{figure}
	\centering
	\includegraphics[width=1\linewidth]{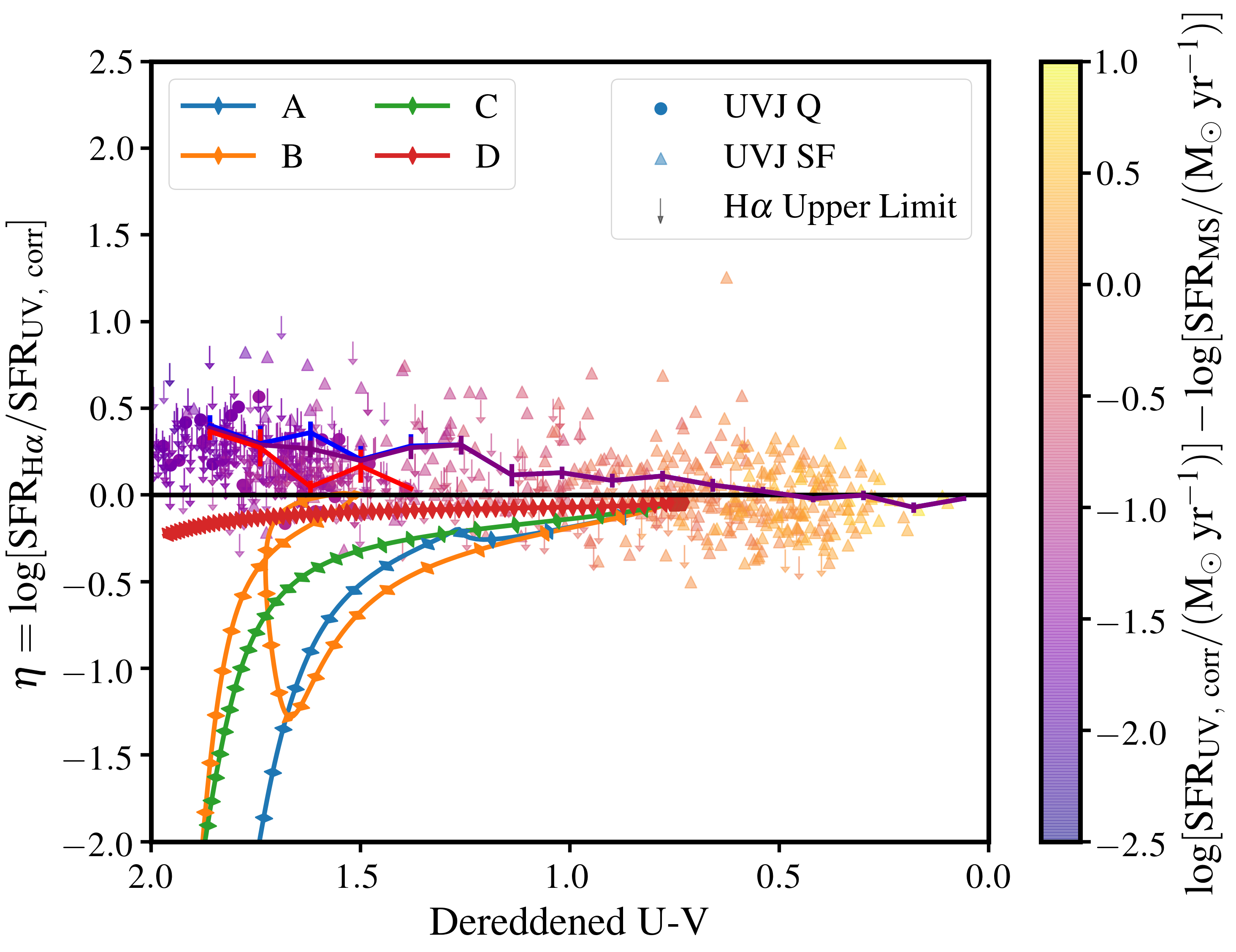}
	\caption{The \ha{}-to-UV ratio as a function of dereddened $U-V$ color. The points are color-coded by the UV SFR. As with Figure~\ref{fig:ratios}, star-forming galaxies have $\eta$ values close to $0$, but the distribution of \ha{} SFRs among red galaxies implies a wide range of $\eta$ values. As with Figure~\ref{fig:ratios}, ticked lines represent the evolution of our models through this space, with ticks every $75$~Myr.}
	\label{fig:ratioscolor}
\end{figure}

\begin{figure}
	\centering
	\includegraphics[width=1\linewidth]{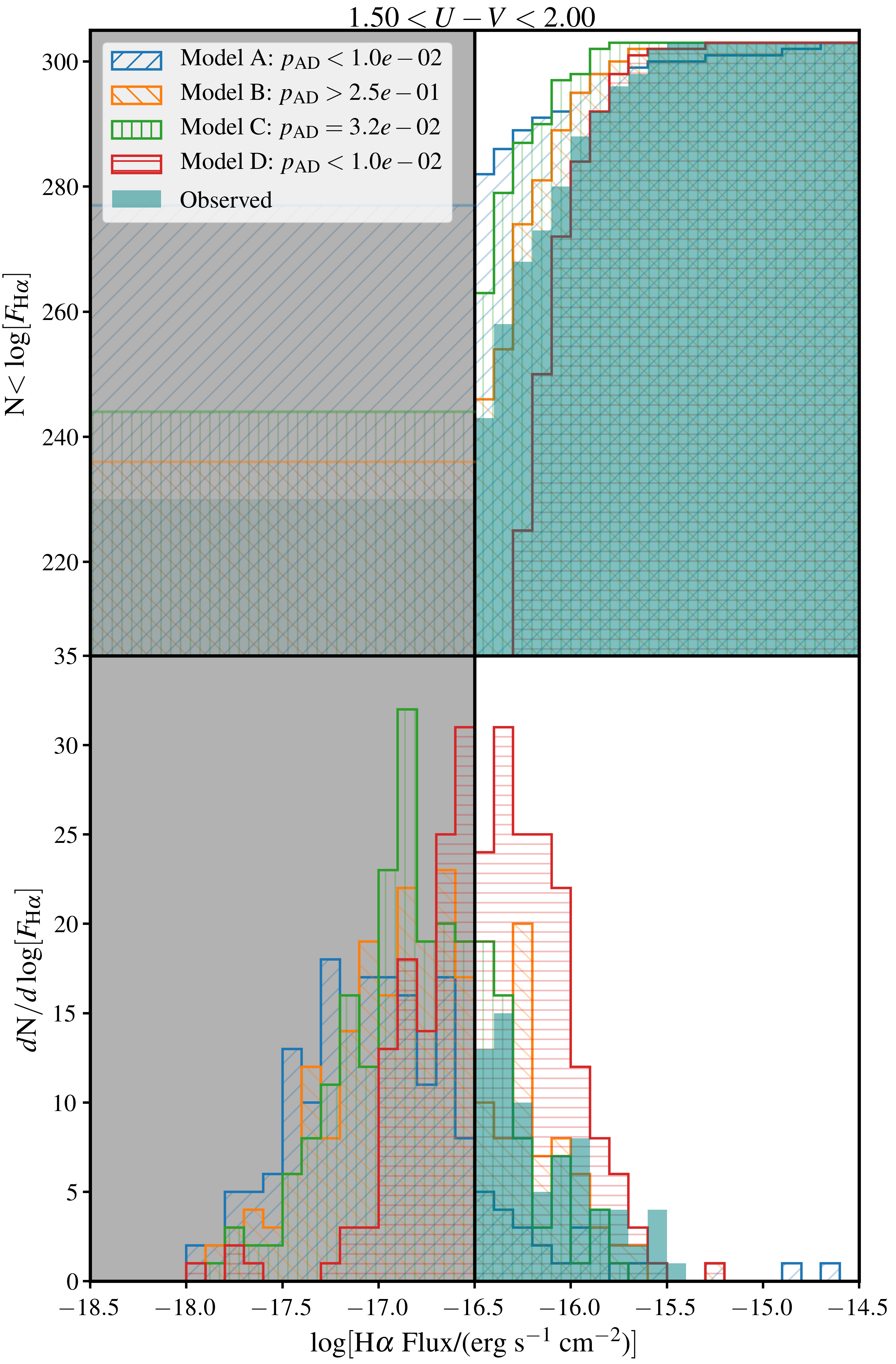}
	\caption{{\bf Bottom:} We compare the observed \ha{} flux distribution with our models for systems with dereddened $U-V$ colors between $1.5$ and $2$. {\bf Top:} The corresponding cumulative distribution, with the legend showing the results of Anderson-Darling tests between the models and data. Again, bursty model B best matches the observed \ha{} distribution. As with Figure~\ref{fig:modelflux}, the shaded region below $F_{{\rm H}\alpha}=10^{-16.5}$~erg~s$^{-1}$~cm$^{-2}$ highlights objects considered non-detections.}
	\label{fig:modelfluxcolor}
\end{figure}

	As an independent test of the SFHs of $z\sim1$ galaxies, we compare the $\eta$ values with rest-frame, de-reddened $U-V$ colors in Figure~\ref{fig:ratioscolor}. Again, we find that blue star-forming galaxies have $\eta$ values close to $0$, whereas red quiescent galaxies have a wide range of $\eta$ values.
	In Figure~\ref{fig:modelfluxcolor}, we show the \ha{} distributions produced through the same method as Figure~\ref{fig:modelflux}, but with $U-V$ color in place of UV SFR. For objects with $1.5<U-V<2$, where $88\%$ of objects more than $1$~dex below the main sequence live, the \ha{} distribution is again most consistent with bursty model B.

	\section{Discussion}
	\label{sec:systematics}
	In this analysis, we find that a bursty SFH is required in order to reproduce the large range in the \ha{}/UV ratios of galaxies with low SFRs. In this section we discuss a few possible origins of this result. First, we discuss the effect of various systematics related to measuring/interpreting the \ha{} and UV SFRs of galaxies in our sample. Next, we discuss possible physical mechanisms driving the range of $\eta$ values observed in our sample.
	\subsection{Systematics}
	\subsubsection{Emission from Active Galactic Nuclei}
	While we account for X-ray AGN in our analysis, it is possible that low-luminosity or obscured AGN contribute to the \ha{} emission of these sources. Indeed \cite{belli2017} have found that a number of $UVJ$-quiescent objects with \ha{} emission have \ntwo{}/\ha{} ratios consistent with an AGN. However, for a wide range of UV luminosities, the inferred \ha{} and UV SFRs are in agreement (see Fig.~\ref{fig:ratios}), so there is no reason to believe that low-level AGN present a significant bias in our modeling. Additionally, X-ray non-detected AGN are not expected to represent a substantial fraction of our sample, further suggesting that the presence of an AGN does not significantly affect our conclusions.
	
	\subsubsection{Emission from Evolved Stars}
	\label{sec:uvevolved}
	UV emission from evolved stars is apparent in local elliptical galaxies \citep{dorman1995}. Studies of the spectra and colors of these galaxies suggest that this emission is primarily due to post-AGB and Blue-Horizontal-Branch stars \citep{yan2018}. This emission is nominally included in the {\sc pyfsps} models, but significant uncertainties remain in our understanding of this phase of stellar-evolution, so our models may be underestimating the UV luminosity of evolved stellar populations. However, this uncertainty does not affect our conclusions: if evolved stars contribute more UV emission than our models, our models should move down and to the right in Figure~\ref{fig:ratios} \emph{except for during a burst}, during which UV emission is dominated by young stars and the contribution from evolved stars is negligible. In this case, smooth models would be a worse fit to the $\eta$ distribution, whereas bursty models would better fit the distribution.

	\subsubsection{Extra Extinction Around H{\sc ii} Regions}
	A significant source of uncertainty regarding this analysis is the amount of extra extinction around H{\sc ii} regions.
	Because the relationship between stellar and nebular extinction solely affects the \ha{} SFR, adjusting this ratio would directly alter our results.
	However, while this relationship is important to ensure that both SFR measures agree for star-forming galaxies, the median \ha{} extinction value for quiescent objects is only $0.3$, so extinction does not play a very important role in calculating the SFRs of the low-SFR population.
	If we adopt the slightly higher nebular-to-continuum extinction ratio (known as the $f$ factor) from \cite{calzetti2000}, model C fits the observations slightly better, but our model B remains the best fitting model. Adopting a lower nebular-to-continuum ratio, as suggested by some recent high-$z$ studies \citep[for ex.~][]{puglisi2016,broussard2019}, improves the fit of model A and weakens the fit of model B. However, using a lower ratio overestimates $\eta$ for systems on the star-forming main sequence:  adopting an $f$ factor of $1$ changes the median $\eta$ value among objects within $0.3$~dex of the main sequence from $-0.06$ to $-0.39$. The suggested variation in $f$ factor at higher redshift is driven by dusty objects with high SFRs \citep{reddy2015}, whereas our sample has sSFRs more similar to low-$z$ objects. Observations of local galaxies suggest that low-sSFR galaxies that have Calzetti-like $f$ factors, whereas high sSFR galaxies (like those at high $z$) have lower $f$ factors \citep{battisti2016}. Additionally, the agreement between Balmer-decrement-based measurements and our $A_{{\rm H}\alpha}$ measurements, as well as the fact that $\eta$ is not correlated with extinction among our sample, suggests that this is not a significant issue (see Fig.~\ref{fig:irvsuv}).

	\subsubsection{Contamination from \normalfont{[N{\sc ii}]}}
	Contamination from nearby \ntwo{} represents a non-negligible contribution to the observed flux (and inferred $\eta$ values). If the \ntwo{}/\ha{} ratio is substantially higher among our sample, the inferred $\eta$ values could be overestimated.
	Indeed, \ntwo{} emission from high-$z$ galaxies appears higher than expected given their [O{\sc III}]/H$\beta$ ratios (\citealt{steidel2014,masters2014,jones2015,shapley2015}, but see \citealt{sanders2018}).
	Increasing the \ntwo{}/\ha{} ratio by $0.37$~dex increases the predicted flux distribution, bringing models A and C in better agreement with observations. The \ntwo{} offset appears to be constant with SFR \citep{strom2017}, which would induce a roughly constant shift in $\eta$ for all SFRs, not a preferentially lower $\eta$ for low-SFR galaxies in particular, as would be necessary for our measurements to match a smoothly declining SFH. Furthermore, there is no residual trend between $\eta$ and stellar mass, as one would expect if this metallicity-dependent effect was important.

	\subsection{Physical Mechanisms}
	\subsubsection{Initial Mass Function}
	As \ha{} emission primarily is dominated by stars with $>8~$\msun{}, whereas UV emission originates from stars with $>4~$\msun{}, the ratio of the two SFR measures is sensitive to the intial mass function (IMF) of the stellar population. In particular, \cite{fumagalli2011} and \cite{dasilva2014} have suggested that stochastic sampling of the initial mass function, both due to limited mass and time resolution at low SFRs may be responsible for variation in \ha{}/UV SFRs. To test the impact of stochastic IMF sampling on our model, we utilize the {\sc slug} code \citep{krumholz2015}. Contrary to traditional stellar-population synthesis codes which integrate a given IMF to a certain mass regardless of the overall SFR, this code directly and stochastically samples the IMF to generate stellar populations. However, the dispersion in the \ha{} to UV ratio for a model with an SFR of $0.1$~\mperyr{} (corresponding to $\Delta$MS$\sim-2$) is only $0.1$~dex (with the fraction of stars formed in clusters set to $1$), not enough to explain the large dispersion in $\eta$ values.

	Alternatively, a very top-heavy IMF could result in systematically higher $\eta$ values compared with the Chabrier values. We test this hypothesis using the \cite{vandokkum2008} parameterization of the IMF, with $M_c$ set to $1.5$ (at the high end of what is observed). In this case, UV and \ha{} emission decrease at similar rates, even for the fast quenching models, such that no model is able to reproduce the large number of galaxies with low $\eta$ values.
	The primary effect of adopting a more-bottom heavy IMF (as suggested by some observations of nearby massive ellipticals \citealt{vandokkum2017}) is a decrease in UV and \ha{} emission at a given SFR. Still, the large dispersion in \ha{} SFRs for galaxies with low UV SFRs cannot be reproduced by any smooth quenching model and is best reproduced by model B.
	Similarly, an integrated galactic-IMF (IGIMF), in which stars are formed primarily in clusters \citep{weidner2005,weidner2011}, results in more top-heavy for galaxies with lower SFRs. Following \cite{pflammaltenburg2007}, we adjust the \ha{}-to-SFR ratio for their {\it Minimal-1} and {\it Standard} models. However, for both models, the variation of the \ha{} luminosity with SFR is not sufficient to explain the observed variation and no SFH model matches the observed distribution with a  $p$ value higher than $0.02$.

	\subsubsection{Minor Mergers}
	Within massive galaxies, bursty star formation is often though of as due to minor mergers or interactions \citep{mihos1994,somerville2008}. For $M_*=10^{10}$~\msun{} galaxies at $z\sim1$, the major merger rate is $10^{-4}$~Mpc$^{-3}$~Gyr$^{-1}$ \citep{duncan2019}, corresponding to $\sim10\%$ per galaxy per Gyr. Assuming that minor merger rate is a factor of $10$ higher than the major merger rate \citep{rodriguez-gomez2015}, we would minor mergers to be common among our sample. This suggests that minor mergers driving bursty star formation could explain the observed burstiness in our population.
	
	\section{Conclusions}
	\label{sec:conclusions}
	Using data from the \dhst{} survey, we analyze \ha{} emission within $\sim800$ massive galaxies at $z\sim1$, focusing on galaxies undergoing the transition between star-forming and quiescence to better understand the process of quenching in these galaxies. Our conclusions are as follows.
	\begin{itemize}
	\item In contrast with expectations, we find evidence of \ha{} emission for galaxies down to the lowest levels of UV SFR present in our sample, including $11\%$ of systems identified as quiescent through the $UVJ$ diagram.
	\item There is a large dispersion ($\sim0.7$~dex) in the ratio between \ha{} and UV SFRs for galaxies with low UV SFRs. Even after accounting for the expected emission from AGN and evolved stars, this large range is inconsistent with a smoothly declining star-formation history.
	\item The observed variation in \ha{}-to-UV SFRs among massive galaxies in the process of quenching implies that quenching at $z\sim1$ is not characterized by a continuous decline in SFR. On the contrary, by modeling various bursty and non-bursty star-formation histories, we show that, bursty star formation continues as the SFR declines.
	\end{itemize}

	Our analysis has been limited to high-mass systems due to the limited S/N of lower mass systems, but given that they have bursty star formation when they are star forming, an analysis of the \ha{} emission in low-mass galaxies transitioning to quiescence would be particularly valuable.

	\section*{Acknowledgments}
We are grateful to the anonymous reviewer, whose suggestions greatly improved this paper.
This research made use of {\texttt{Astropy}}, a community-developed core Python
package for Astronomy \citep{astropy}. Additionally, the Python packages
{\texttt{NumPy}} \citep{numpy}, {\texttt{iPython}} \citep{ipython},
{\texttt{SciPy}} \citep{scipy}, and {\texttt{matplotlib}} \citep{matplotlib}
were utilized for the majority of our data analysis and presentation.

	\appendix
	\section{Other Star-formation Histories}
	While it is beyond the scope of this work to evaluate all possible bursty SFHs, we explore the effects of varying the burst timescale, delay time, and strength in this appendix. In Table~\ref{tab:modeltable2}, we describe $4$ SFHs that we explore beyond the $4$ described in our primary analysis in the form of
	\begin{equation}
	{\rm SFR}={\rm SFR_0}e^{-(t-t_q)/\tau_0}+\delta e^{{|t-t_q-t_1|}/\tau_{\rm burst}},
	\end{equation}
	with $\delta$ representing the burst amplitude, $t_q$ representing the time of quenching, $t_1$ representing the burst delay time, and $\tau_{\rm burst}$ representing the exponential timescale of the burst. 
	
\begin{figure}
		\centering
		\includegraphics[width=1\linewidth]{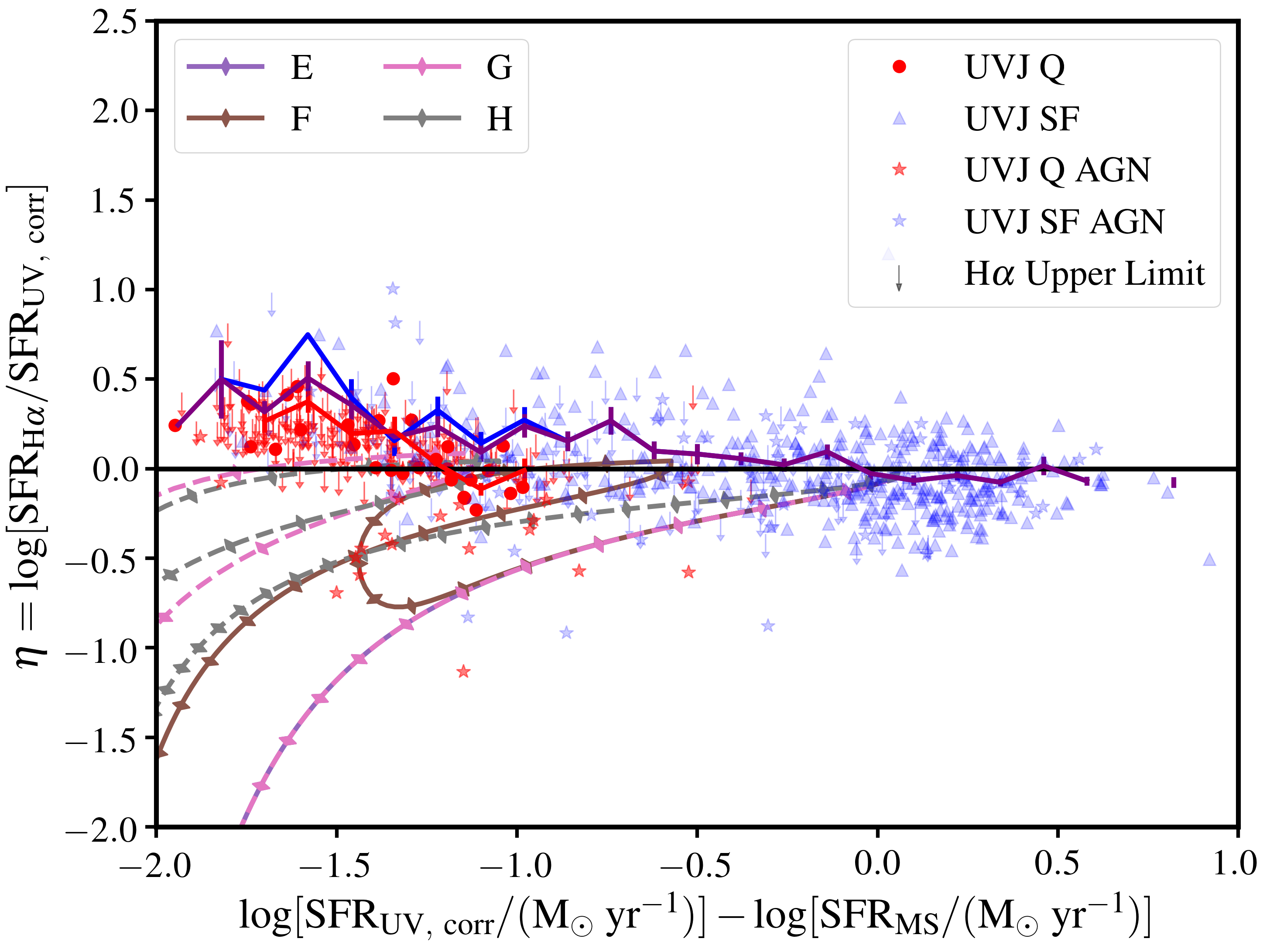}
		\caption{The evolution of $\eta$ with $\Delta$MS for SFH  models E--H.}
		\label{fig:appendix1}
	\end{figure}

\begin{figure}
		\centering
		\includegraphics[width=1\linewidth]{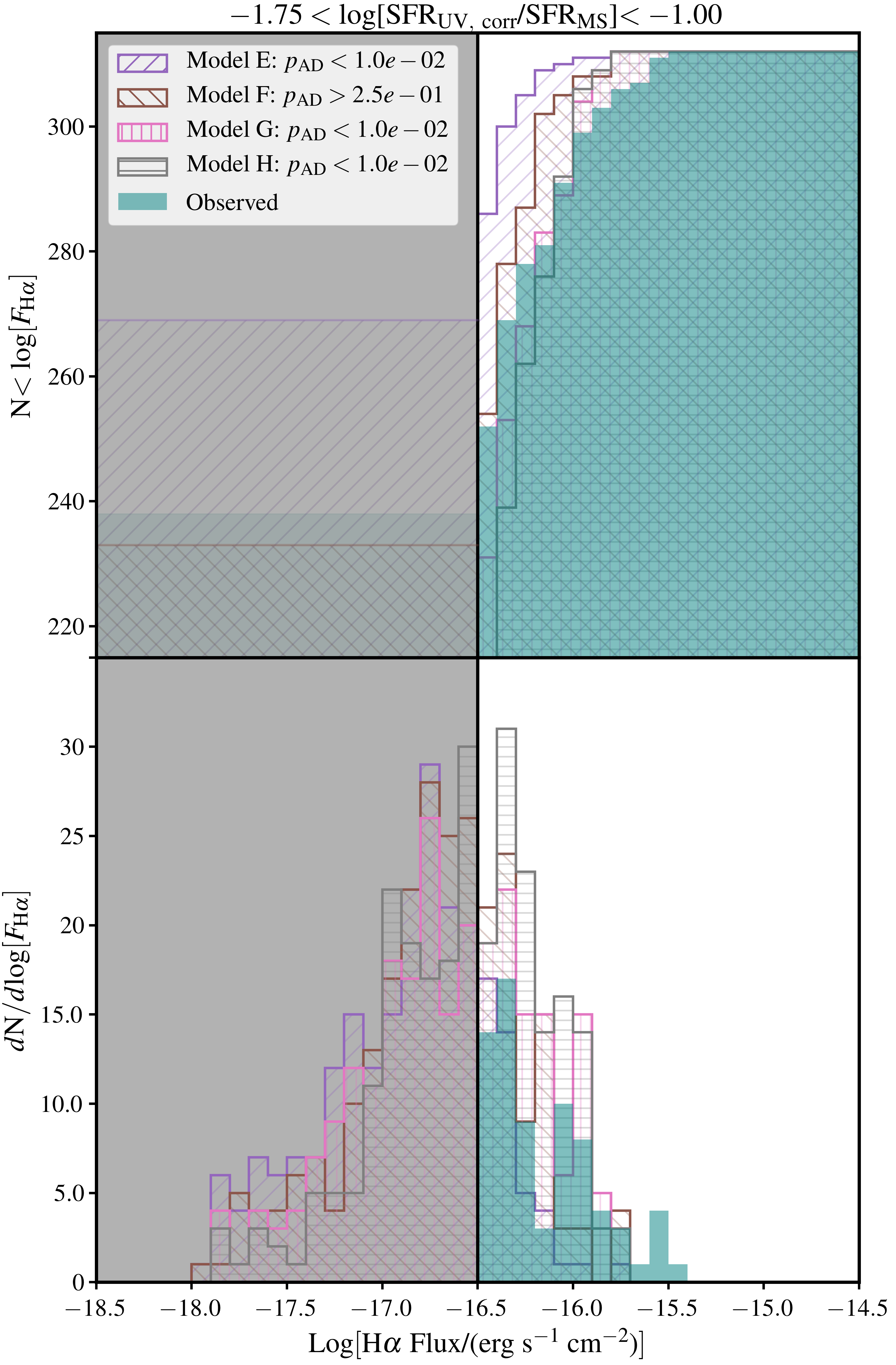}
		\caption{The \ha{} flux distribution for SFH models E--H. Even considering the additional models here, it is clear that bursty models fit the observations better than smooth models overall.}
		\label{fig:appendix2}
	\end{figure}

	Figure~\ref{fig:appendix1} shows the relationship between $\eta$ and $\Delta$MS for these SFHs in comparison with our observations. The most important variable is the burst delay time: models with high $t_1$ values are able to reach lower SFR values before bursting. As shown in Figure~\ref{fig:appendix2}, bursty model F is preferred to any smoothly declining model.
	
	We also consider linearly increasing, delayed, and inverted tau models in the form of:
	\begin{equation}
	\begin{array}{lcr}
	{\rm SFR}(t)=
	\begin{cases}
	{\rm SFR_0} (t/t_1) &  t<t_1 \\
	{\rm SFR_0} e^{-t/\tau_0} + \delta e^{-|t-t_1|/\tau_{\rm burst}} & t>t_1,
	\end{cases}
	\end{array}
	\label{eqn:sfhlinear}
	\end{equation}
	\begin{equation}
	{\rm SFR}(t)={\rm SFR_0} (t-t_1)e^{-(t-t_1)/\tau_0} + \delta e^{-(t-t_1)/\tau_{\rm burst}},
	\label{eqn:sfhdelayed}
	\end{equation}
	and
	\begin{equation}
	\begin{array}{lcr}
	{\rm SFR}(t)=
	\begin{cases}
	{\rm SFR_0} e^{t/\tau_1} &  t<t_1 \\
	{\rm SFR_0} e^{-t/\tau_0} + \delta e^{-|t-t_1|/\tau_{\rm burst}} &  t>t_1\;,
	\end{cases}
	\end{array}
	\label{eqn:sfhinverted}
	\end{equation}
	with SFR$_0$, $\delta$, $t_1$, $\tau_0$, and $\tau_{\rm burst}$ values as in models A, B, C, and D. The parameters describing these models as well as the results of our comparison of these models with observations are found in Table~\ref{tab:modeltable2}. 
		For the delayed model (equation~\ref{eqn:sfhdelayed}), no model accurately reproduces the \ha{} flux distribution for objects between $1$ and $1.75$~dex below the main sequence.
		For the inverse model (equation~\ref{eqn:sfhinverted}), bursty model~B reproduces the \ha{} flux distribution, whereas smooth model~C does not.
		Lastly, for the linearly increasing model (equation~\ref{eqn:sfhlinear}), both model~B and C reproduce the observed \ha{} flux distribution for objects between $1$ and $1.75$~dex below the main sequence. Model~B fits the \ha{} distribution for objects with dust-corrected $U-V$ colors between $1$ and $2$ and model~C does not, however.
		In summary, regardless of the general form of the star-formation history adopted in our models, a bursty star-formation history better fits the observed \ha{} fluxes compared with a smoothly-declining star-formation history.

			\begin{table*}
		\centering
		\begin{tabular}{|c|c|c|c|c|c|c|c|c}
			\hline 
			Model & SFH Equation & $\delta$ (\mperyr{}) & $t_1$ (Gyr) & $\tau_0$ (Myr) & $\tau$ (Myr) & $p$ value ($\Delta$MS) & $p$ value (dereddened $U-V$)  \\ 
			\hline
			E \label{model:earlyhigh} & \ref{eqn:sfh}& 0 & -- & 100 & -- & $<0.01$ & $<0.01$\\ 
			\hline 
			F \label{model:latehigh} & \ref{eqn:sfh}& 3 & 1 & 100 & 100 & $>0.25$& $<0.01$\\ 
			\hline 
			G \label{model:earlyshort} & \ref{eqn:sfh}& 1 & 1 & 200 & 200& $<0.01$& $<0.01$ \\ 
			\hline 
			H \label{model:lateshort} & \ref{eqn:sfh}& 3 & 1 & 200 &200 & $<0.01$& $<0.01$\\ 
			\hline 
			A Increasing & \ref{eqn:sfhlinear}& 1 & 0.3 & 100 &100 & $<0.01$& $<0.01$\\ 
			\hline 
			B Increasing & \ref{eqn:sfhlinear}& 1 & 1.25 & 100 &100 & $>0.25$& $0.11$\\ 
			\hline 
			C Increasing & \ref{eqn:sfhlinear}& 0 & -- & 200 & -- & $0.16$& $<0.01$\\ 
			\hline 
			D Increasing& \ref{eqn:sfhlinear} & 0 & -- & 1000 & -- & $<0.01$& $<0.01$\\ 
			\hline 
			A Delayed & \ref{eqn:sfhdelayed}& 1 & 0.3 & 100 &100 & $<0.01$ & $<0.01$\\ 
			\hline 
			B Delayed & \ref{eqn:sfhdelayed}& 1 & 1.25 & 100 &100 & $<0.01$& $<0.01$\\ 
			\hline 
			C Delayed& \ref{eqn:sfhdelayed} & 0 & -- & 200 & -- & $<0.01$& $<0.01$\\ 
			\hline 
			D Delayed & \ref{eqn:sfhdelayed} & 0 & -- & 1000 & -- & $<0.01$& $<0.01$\\ 
			\hline 
			A Inverse & \ref{eqn:sfhinverted}& 1 & 0.3 & 100 &100 & $0.035$ & $<0.01$\\ 
			\hline 
			B Inverse& \ref{eqn:sfhinverted} & 1 & 1.25 & 100 &100 & $>0.25$ & $>0.25$\\ 
			\hline 
			C Inverse & \ref{eqn:sfhinverted}& 0 & -- & 200 & -- & $<0.01$& $0.032$\\ 
			\hline 
			D Inverse & \ref{eqn:sfhinverted}& 0 & -- & 1000 & -- & $<0.01$& $<0.01$\\ 
			\hline 
			
		\end{tabular} 
		\caption{The parameters of the model SFHs used in our extended analysis, along with the $p$ value of an Anderson-Darling test comparing model systems with observations. Column 7 shows the comparison with objects between $1$ and $1.75$ dex below the main sequence and column 8 shows the $p$ value resulting from the comparison with objects with dereddened U-V colors between $1.5$ and $2$. }
		\label{tab:modeltable2}
	\end{table*}

	\bibliography{quenchburstbib}
\end{document}